\documentclass[twocolumn]{aastex63}

\usepackage[T1]{fontenc}  
\usepackage{amsmath}      
\usepackage{mathtools}    
\usepackage{hyperref}     

\hypersetup{colorlinks=true,linkcolor=black,citecolor=black,filecolor=black,urlcolor=black}

\expandafter\def\csname editcolor1\endcsname{blue}

\makeatletter
\newcommand{\extref}[1]{\textup{\tagform@{#1}}}
\makeatother

\DeclarePairedDelimiter{\paren}{\lparen}{\rparen}
\DeclarePairedDelimiter{\unit}{\lbrack}{\rbrack}

\newcommand{\xmin}{x_\mathrm{min}}
\newcommand{\xmax}{x_\mathrm{max}}
\newcommand{\rhor}{r_\mathrm{hor}}
\newcommand{\pgas}{p_\mathrm{gas}}
\newcommand{\pmag}{p_\mathrm{mag}}
\newcommand{\tti}{T_\mathrm{i}}
\newcommand{\tte}{T_\mathrm{e}}
\newcommand{\rhigh}{R_\mathrm{high}}
\newcommand{\rlow}{R_\mathrm{low}}
\newcommand{\mdotedd}{\dot{M}_\mathrm{Edd}}
\newcommand{\sigmat}{\sigma_\mathrm{T}}
\newcommand{\rrring}{R_\mathrm{ring}}
\newcommand{\rrshadow}{R_\mathrm{shadow}}
\newcommand{\eshadow}{e_\mathrm{shadow}}
\newcommand{\rring}{r_\mathrm{ring}}

\newcommand{\ghz}{\mathrm{GHz}}
\newcommand{\msun}{M_\odot}
\newcommand{\mprot}{m_\mathrm{p}}
\newcommand{\pc}{\mathrm{Mpc}}
\newcommand{\mpc}{\mathrm{Mpc}}
\newcommand{\kpc}{\mathrm{kpc}}
\newcommand{\jy}{\mathrm{Jy}}
\newcommand{\gcc}{\mathrm{g\ cm^{-3}}}
\newcommand{\seconds}{\mathrm{s}}
\newcommand{\minutes}{\mathrm{min}}
\newcommand{\hours}{\mathrm{hr}}
\newcommand{\years}{\mathrm{yr}}
\newcommand{\mdot}{\dot{M}}
\newcommand{\msunyr}{\msun\ \years^{-1}}
\newcommand{\muas}{\mathrm{\mu as}}

\newcommand{\athena}{\texttt{Athena++}}
\newcommand{\grtrans}{\texttt{grtrans}}

\defcitealias{EHT2019a}{EHT~I}
\defcitealias{EHT2019b}{EHT~II}
\defcitealias{EHT2019d}{EHT~IV}
\defcitealias{EHT2019e}{EHT~V}
\defcitealias{EHT2019f}{EHT~VI}

\shorttitle{Effects of Tilt on Images}
\shortauthors{White, Dexter, Blaes, and Quataert}

\begin{document}

\title{The Effects of Tilt on the Images of Black Hole Accretion Flows}
\author{Christopher~J.~White}
\affiliation{Kavli Institute for Theoretical Physics, University of California Santa Barbara, Kohn Hall, Santa Barbara, CA 93107, USA}
\author{Jason~Dexter}
\affiliation{JILA and Department of Astrophysical and Planetary Sciences, University of Colorado Boulder, Boulder, CO 80309, USA}
\author{Omer~Blaes}
\affiliation{Department of Physics, University of California Santa Barbara, Broida Hall, Santa Barbara, CA 93106, USA}
\author[0000-0001-9185-5044]{Eliot~Quataert}
\affiliation{Department of Astronomy, University of California Berkeley, 501 Campbell Hall, Berkeley, CA 94720, USA}

\begin{abstract}
  We analyze two 3D general-relativistic magnetohydrodynamic accretion simulations in the context of how they would manifest in Event Horizon Telescope (EHT) observations of supermassive black holes. The two simulations differ only in whether the initial angular momentum of the plasma is aligned with the rapid ($a = 0.9$) spin of the black hole. Both have low net magnetic flux. Ray tracing is employed to generate resolved images of the synchrotron emission. When using parameters appropriate for Sgr~A* and assuming a viewing angle aligned with the black hole spin, we find the most prominent difference is that the central shadow in the image is noticeably eccentric in tilted models, with the ring of emission relatively unchanged. Applying this procedure to M87 with a viewing angle based on the large-scale jet, we find that adding tilt increases the angular size of the ring for fixed black hole mass and distance, while at the same time increasing the number of bright spots in the image. Our findings illustrate observable features that can distinguish tilted from aligned flows. They also show that tilted models can be viable for M87, and that not accounting for tilt can bias inferences of physical parameters. Future modeling of horizon-scale observations should account for potential angular momentum misalignment, which is likely generic at the low accretion rates appropriate for EHT targets.
\end{abstract}

\section{Introduction}
\label{sec:introduction}

The Event Horizon Telescope (EHT) can now produce resolved images of black hole accretion flows. There is a robust observation of a ring of light around the supermassive black hole in M87 \citep[\citetalias{EHT2019a}]{EHT2019a}, with observations of polarization, as well as of Sgr~A*, coming soon. Additionally, we expect the near future to feature even more detailed observations of this sort, with additional interferometric baselines, more sensitivity, and alternate frequencies \citep[\citetalias{EHT2019b}]{EHT2019b}.

These observational capabilities allow unprecedented direct comparison to general relativistic (GR) simulations of black hole accretion flows. From the perspective of magnetohydrodynamics (MHD), such flows are generally characterized by the black hole mass and spin, the density scale of the accreting matter, the geometrical thickness of the flow, and the strength and configuration of the magnetic field. The flow is expected to be geometrically thick for both M87 and Sgr~A* \citep{Narayan1998,Blandford1999,Yuan2014}, and the effects of varying spin, disk magnetization, and to a limited extent electron temperature prescription are considered in a library of simulation images in \citet[\citetalias{EHT2019e}, \citetalias{EHT2019f}]{EHT2019e,EHT2019f}.

However there exists another important parameter that can have a significant impact on the appearance of an accretion flow:\ the misalignment between the black hole spin and the angular momentum of the infalling matter. In the case of low-luminosity active galactic nuclei (AGN) such as M87 and Sgr~A*, we do not expect this matter to be aligned with the spin at large radii. The gas matter angular momentum cannot be quickly torqued into alignment with the black hole spin given the geometrically thick disks present at low accretion rates ($\mdot$), nor can a black hole of mass $M$ be torqued into alignment with the disk via accretion of angular momentum except on timescales comparable to $M / \mdot$. Early GR simulations of such flows were performed by \citet{Fragile2005} and \citet{Fragile2007}, where they report a number of qualitative differences between aligned and tilted flows. The disks become warped and twisted by differential Lense--Thirring precession, and a pair of standing shocks can develop \citep{Fragile2008,Generozov2014}.

These standing shocks in particular may dissipate a large amount of kinetic energy locally, heating electrons and causing non-axisymmetric emission. From the parameter survey of \citet{White2019}, we expect this dissipation in standing shocks to be comparable to the dissipation arising from turbulence and magnetic reconnection throughout the rest of the accretion disk whenever the dimensionless spin is sufficiently large, $a \gtrsim 0.9$, and the flow is sufficiently tilted, with an inclination $i \gtrsim 8^\circ$. Indeed, $i = 15^\circ$ models of Sgr~A* have been shown to differ from aligned models in $230\ \ghz$ images \citep{Dexter2013}.

Here we investigate these effects further, contrasting the appearances of two GRMHD models, one aligned and one tilted, in different contexts. \S\ref{sec:numerics} describes the simulations and the subsequent ray tracing used to produce images comparable to what is seen by the EHT. In \S\ref{sec:analysis} we illustrate the effects of tilt in three different settings. First we consider images as seen when looking along the spin axis, where the effects of the standing shocks are most intuitive, modeling Sgr~A* (whose orientation is currently not well constrained). We then consider the same system but viewed $45^\circ$ off the spin axis. Finally, we analyze the case of M87, where the orientation can be inferred from observations at large scales. Here tilted disk models produce asymmetric ring (crescent) morphologies that appear compatible with the EHT image.

We will often refer to ``rings'' and ``shadows,'' by which we simply mean the structures in images at current EHT resolution. These general terms will not necessarily refer to the photon ring or black hole shadow, which are fixed properties of the spacetime and do not depend on the structure of the surrounding emission as long as it extends inside of the black hole photon orbit. Our goal is to provide a sense of what observable properties would signify the presence or absence of angular momentum misalignment. Our results on shadow shape and size have implications for GR tests based on the properties of EHT images, unless the direct accretion flow emission can be separated from that corresponding to the photon ring itself, as we discuss in \S\ref{sec:implications}.

Throughout this work, quantities with units omitted are taken to be in geometric units appropriate for GRMHD simulations around a black hole of mass $M$:\ length in units of $GM/c^2$, time in units of $GM/c^3$, and density scaled arbitrarily.

\section{Numerical Procedure}
\label{sec:numerics}

\subsection{GRMHD Simulations}
\label{sec:numerics:simulations}

We use the GRMHD code \athena{} \citep{White2016} to evolve two similar accretion flows around a black hole with spin $a = 0.9$, one aligned with the black hole spin and the other tilted. In both cases we use spherical Kerr--Schild coordinates $(t, r, \theta, \phi)$ with a statically refined grid. The root grid has $56$ cells geometrically spaced in radius from $r \approx 0.926\ \rhor$ (the horizon is at $\rhor \approx 1.44$) to $r = 100$, $32$ cells uniformly spaced in $\theta$ from pole to pole, and $44$ cells uniformly spaced in $\phi$ from $0$ to $2\pi$. Three nested levels of mesh refinement are added, each doubling resolution in all three dimensions. The highest effective resolution of $448 \times 256 \times 352$ ($239$ cells per decade in radius) is achieved everywhere within $50.625^\circ$ of the midplane.

The aligned simulation is initialized with a hydrostatic torus according to the prescription of \citet{Fishbone1976}, with inner edge $r = 15$, pressure maximum at $r = 25$, adiabatic index $\Gamma = 4/3$, and peak density $\rho = 1$. A poloidal magnetic field is added using the vector potential
\begin{multline}
  A_\phi \propto \paren[\big]{\max(\pgas - 10^{-8}, 0)}^{1/2} r^2 \sin\theta \\
  \times \sin\paren[\big]{\pi L(r; 16, 34)} \sin\paren[\big]{\pi L(\theta; 70^\circ, 110^\circ)},
\end{multline}
with $A_r, A_\theta = 0$. Here $L(x; \xmin, \xmax)$ is the linear ramp function that runs from $0$ for $x \leq \xmin$ to $1$ for $x \geq \xmax$.

The tilted simulation applies the \citeauthor{Fishbone1976} solution to the coordinates $(t, r, \theta', \phi')$, with
\begin{subequations} \begin{align}
  \theta' & = \cos^{-1}(\cos i \cos\theta + \sin i \sin\theta \cos\phi), \\
  \phi' & = \tan^{-1}(\sin\theta \sin\phi, \notag \\
  & \quad \qquad -\sin i \cos\theta + \cos i \sin\theta \cos\phi)
\end{align} \end{subequations}
being the standard angles obtained by tilting spherical coordinates by an inclination $i = 24^\circ$ toward $\phi = 0^\circ$. The torus is no longer in exact equilibrium, though the addition of a magnetic field causes this in any event. Here we apply the vector potential
\begin{equation}
  A_{\phi'} \propto \max(\rho - 0.2, 0),
\end{equation}
with $A_r, A_{\theta'} = 0$.

In both cases we have a magnetic field consisting of a single set of nested loops in the poloidal plane. We normalize the fields such that the density-weighted average of plasma $\beta^{-1} \equiv \pmag / \pgas$ is $0.01$.

\begin{figure}
  \centering
  \includegraphics{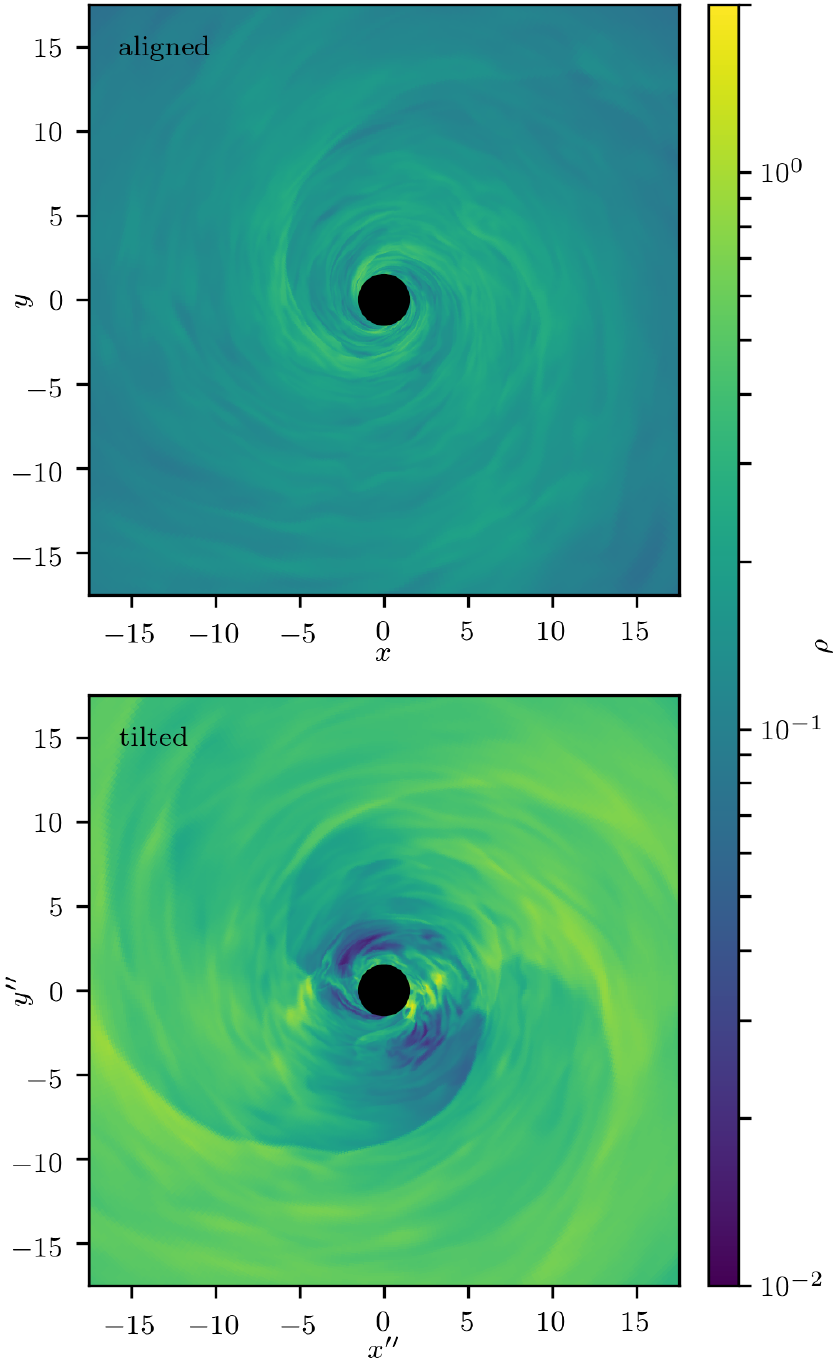}
  \caption{Midplane slices of density at the end of the two simulations. In the tilted case, the midplane is warped and twisted to follow the shell-averaged angular momentum of the disk. There are stronger density contrasts in the tilted case than in the aligned case. \label{fig:slices}}
\end{figure}

Both simulations are run to a time of $t = 10{,}000$. As expected, the structures of the flows differ in the two cases. Figure~\ref{fig:slices} shows midplane slices of density at the end of the simulation. In the tilted case, the surface used for the slicing, denoted with two primes, is warped and twisted in order to be orthogonal to the gas angular momentum at each radius. The angle between the surface normal and the black hole spin direction can exceed the initial inclination, reaching $44^\circ$ at small radii. Using the same density scale, one can see much stronger density contrasts in the tilted case. These result from the standing shocks created in misaligned disks, as explained by \citet{Fragile2008}.

\subsection{Ray Tracing}
\label{sec:numerics:ray_tracing}

We use the GR ray tracing code \grtrans{} \citep{Dexter2009,Dexter2016} to process snapshots from the simulations into images. The camera is located at coordinates $(r_0, \theta_0, \phi_0)$. We set $r_0 = 50$; in all cases there is negligible emission and absorption beyond this radius. A grid of $512^2$ pixels is created in the image plane, with a field of view of $24\ GM/c^2$ on each side. Each ray traced back from the image plane through the simulation is sampled at $1600$ points.

Here we only consider total intensity (Stokes $I$) images (only tracking polarization internally during ray tracing) at $230\ \ghz$ as produced by the emission and absorption of thermal synchrotron radiation along each ray. The scale-free MHD snapshots are given physical units by setting the black hole mass $M$ and fluid density scale $\unit{\rho}$. The latter is adjusted in order to match observed $230\ \ghz$ flux density, fixing the time-averaged mass accretion rate onto the black hole in the process. In all cases the fluid temperature in the simulation is converted to electron temperature according to the same ansatz as in \citet{Moscibrodzka2016}, assuming an ion-to-electron temperature ratio of
\begin{equation}
  \frac{\tti}{\tte} = \frac{\rlow + \rhigh \beta^2}{1 + \beta^2},
\end{equation}
with $\rlow = 1$ and $\rhigh = 10$. This is the same prescription used in \citetalias{EHT2019e}, where $\rlow$ is also kept at $1$ and $\rhigh$ is varied over a range of plausible values from $1$ to $160$. We choose to use a single intermediate $\rhigh$ in order to focus on the effects of accretion flow geometry. While changing this parameter can redistribute intensity within the image, possibly changing the overall size of the observed ring, we do not expect variations in $\rhigh$ to qualitatively change aspects of image morphology that arise due to flow geometry. We note that models developed to capture the effects of turbulent heating may not be appropriate for shock-heated regions in the tilted flows, and that better prescriptions may need to be developed for these cases.

When applying the simulations to a face-on model of Sgr~A*, we choose $M = 4.152 \times 10^6\ \msun$ and place the source at a distance $D = 8.178\ \kpc$, following the inferred values in \citet{Gravity2019}. We seek to have the average flux from our snapshots match $F_\nu = 2.4\ \jy$ \citep{Doeleman2008}, which results in choosing $\unit{\rho} = 1.7 \times 10^{-16}\ \gcc$ in the aligned case and $\unit{\rho} = 4.8 \times 10^{-17}\ \gcc$ in the tilted case. With these scales we can convert the accretion rates in the simulations, averaged from $t = 8000$ to $t = 10{,}000$, to the physical values of $\mdot = 8.4 \times 10^{-9}\ \msunyr$ in the aligned case and $\mdot = 5.1 \times 10^{-9}\ \msunyr$ in the tilted case. Defining an Eddington accretion rate of $\mdotedd = 10 \cdot 4\pi G M \mprot / c \sigmat$, these correspond to $9.1 \times 10^{-8}\ \mdotedd$ and $5.5 \times 10^{-8}\ \mdotedd$. Here we fix the viewing angle $\theta_0 = 177^\circ$ and we choose $\phi_0 = 0^\circ$, though the symmetry of the system means the latter has little effect. The images are rotated such that the south pole (the one pointed toward the camera) has a small projection in the plane of the image at a position angle of $180^\circ$ (toward the bottom), with the flow moving clockwise.

For the inclined model of Sgr~A*, we keep the same mass, distance, and flux. The viewing angle is $\theta_0 = 135^\circ$, $\phi_0 = 0^\circ$. Again the southern (near) pole points to the bottom of the image and the matter is moving clockwise. In this case, the density scale $\unit{\rho}$ is $1.6 \times 10^{-16}\ \gcc$ in the aligned case and $5.6 \times 10^{-17}\ \gcc$ in the tilted case. The accretion rates are $7.7 \times 10^{-9}\ \msunyr$ ($8.4 \times 10^{-8}\ \mdotedd$) and $6.0 \times 10^{-9}\ \msunyr$ ($6.5 \times 10^{-9}\ \mdotedd$), respectively.

With M87 we use the inferred mass $M = 6.5 \times 10^9\ \msun$ and combined distance measurement $D = 16.8\ \mpc$ from \citetalias{EHT2019f} (the distance is derived from the measurements in \citet{Blakeslee2009,Bird2010,Cantiello2018}), as well as the flux $F_\nu = 0.98\ \jy$ \citep{Doeleman2012}, resulting in $\unit{\rho}$ being $4.8 \times 10^{-18}\ \gcc$ and $6.9 \times 10^{-19}\ \gcc$ in the aligned and tilted cases, respectively. This implies physical accretion rates of $\mdot = 5.8 \times 10^{-4}\ \msunyr$ ($4.0 \times 10^{-6}\ \mdotedd$) and $\mdot = 1.8 \times 10^{-4}\ \msunyr$ ($1.2 \times 10^{-6}\ \mdotedd$), respectively. In both cases we use a viewing angle $\theta_0 = 163^\circ$ \citep[agreeing in magnitude with][]{Mertens2016}, fixing $\phi_0 = 0^\circ$. Here we rotate the image such that an approaching jet aligned with the spin axis will have a position angle of $288^\circ$ (toward the right and slightly up), in agreement with the large-scale jet seen in M87 \citep{Walker2018}. The accretion flow is clockwise.

In the process of ray tracing, we can delineate the boundary between rays which trace back into the black hole and those that do not. We will refer to this boundary as the ``geometrical ring,'' also known as the photon ring.

In several cases we will consider a set of $21$ snapshots uniformly sampled in time from $t = 8000$ to $t = 10{,}000$. The separation of $\Delta t = 100$ between snapshots is larger than the correlation time at the small radii of interest in these turbulent accretion flows, and so these samples can be considered independent.

\section{Image Analysis}
\label{sec:analysis}

\subsection{Face-On Shadow Shape in Sgr~A*}
\label{sec:analysis:sgra_03}

First we consider the face-on case. Here we use parameters suitable for Sgr~A*, where the inclinations of both the spin axis and the gas angular momentum to the line of sight are currently poorly constrained. For example, one might expect the system to prefer alignment with the clockwise disk (that is more edge-on than face-on) of stars within $0.3\ \pc$ \citep{Paumard2006,Beloborodov2006}, but at the same time orbital motion in a nearly face-on system is consistent with the near-infrared centroid motion found by the GRAVITY instrument \citep{Gravity2018}. We choose to first model the system with face-on spin, where the connection between shock structure and image features is clearest.

\begin{figure}
  \centering
  \includegraphics{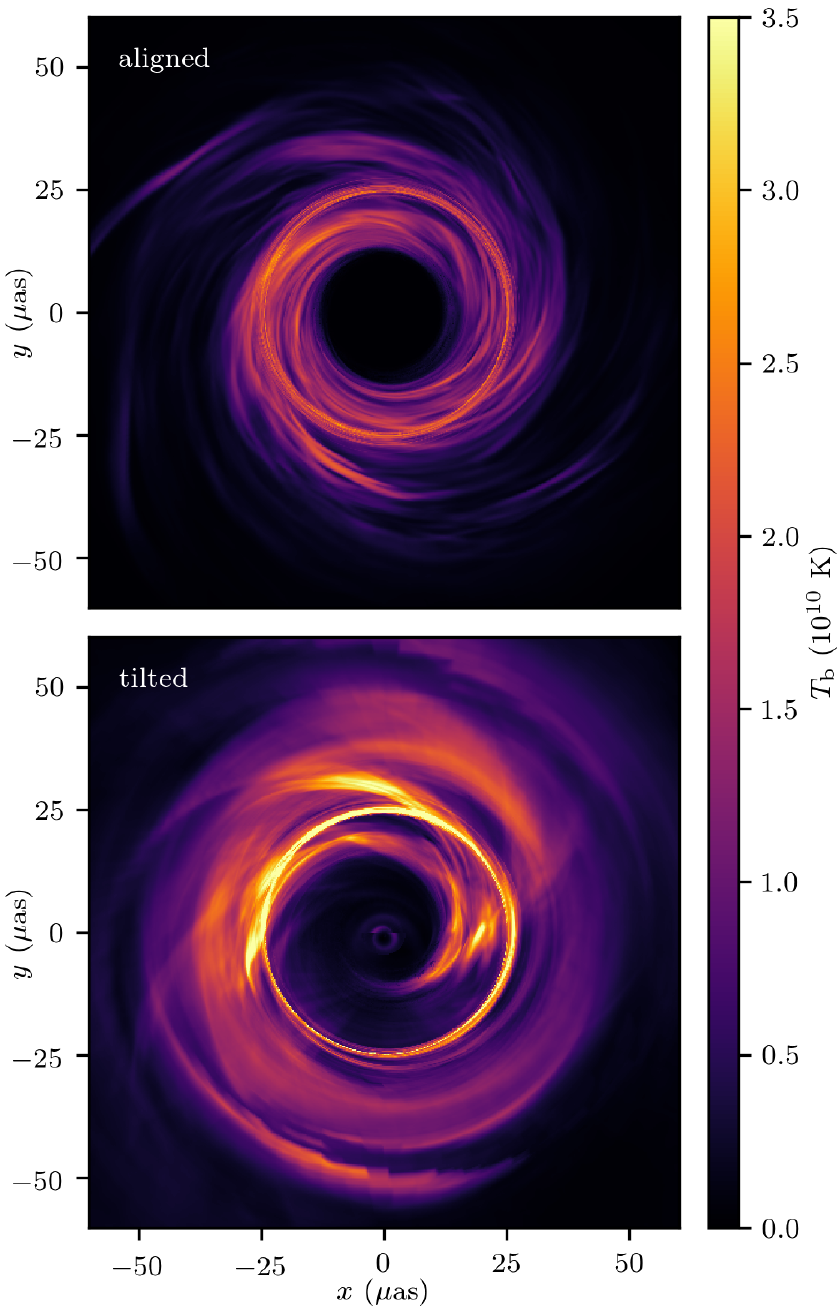}
  \caption{Images created from a snapshot from each simulation, modeled on Sgr~A* assuming a nearly face-on viewing angle. The accretion flow is clockwise in both images, with the southern spin axis pointing toward the camera $3^\circ$ off the line of sight (pointing down when projected onto the image plane). The tilted disk induces the foreground non-axisymmetric shock seen in emission, spiraling clockwise inward just to the right of the center of the image. \label{fig:unblurred_sgra_03}}
\end{figure}

Figure~\ref{fig:unblurred_sgra_03} shows images constructed from high-resolution snapshots at the end of the simulation ($t = 10{,}000$). Here we can clearly see a breaking of axisymmetry caused by the standing shocks in the tilted disk, as well as by Doppler shifts in the parts of the disk moving toward or away from the camera. Importantly, the standing shock seen here is in the foreground (the disk is geometrically thick with the inner part not entirely optically thin, and the other shock in the background is obscured and distorted), enabling it to be seen at small projected radii where there is only an uninterrupted, circular shadow in the aligned case.

The resolution of Figure~\ref{fig:unblurred_sgra_03} is far higher than that of the EHT. We therefore blur the image with a Gaussian kernel with a full width at half-maximum of $20\ \muas$ to see if there remain any features capable of distinguishing tilted from aligned flows. This kernel is appropriate for modeling EHT data in the image plane \citepalias{EHT2019b}. In particular, we expect the bright, nonaxisymmetric feature near the center of the lower panel of Figure~\ref{fig:unblurred_sgra_03} to affect the shadow shape even if the ring remains relatively circular.

\begin{figure*}
  \centering
  \includegraphics{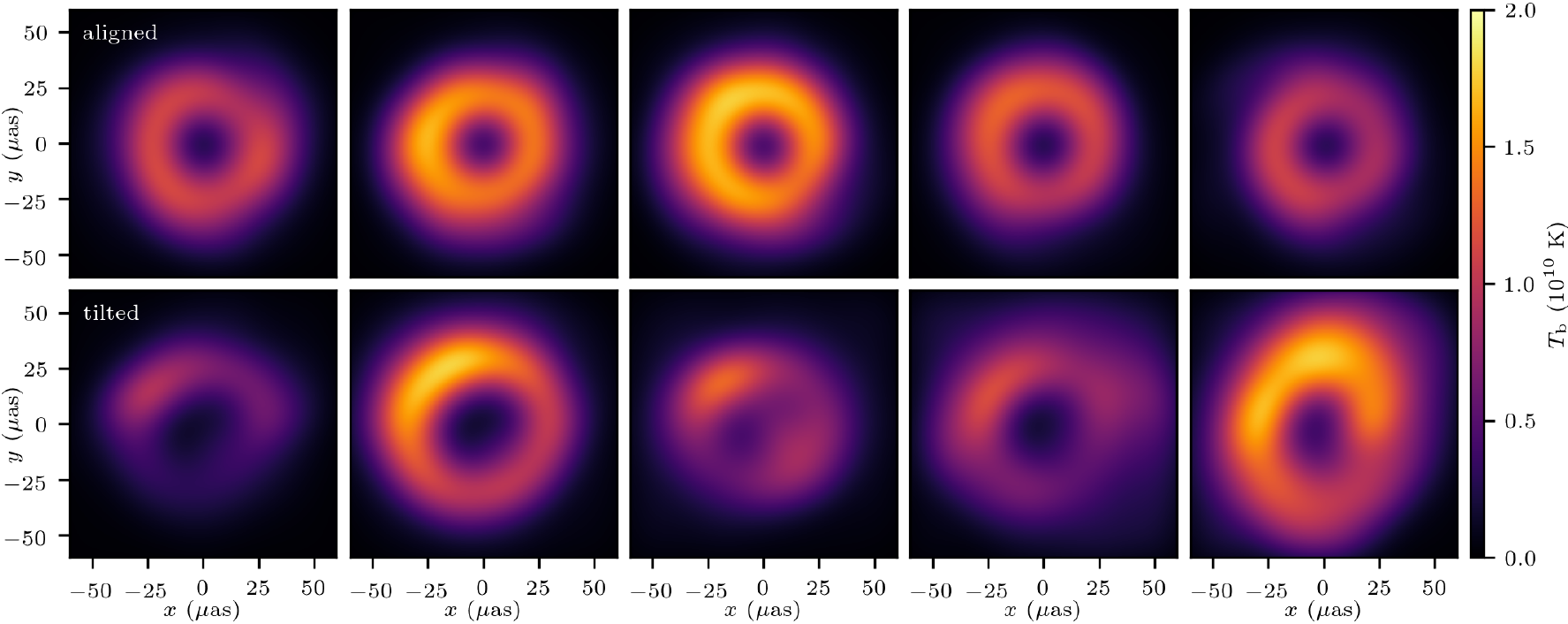}
  \caption{Blurred images modeled on Sgr~A* with a nearly face-on viewing angle for the aligned (top) and tilted (bottom) simulations. The camera position and orientation is the same as in Figure~\ref{fig:unblurred_sgra_03}. The snapshots span a time range of $11\ \hours$. While the bright shock interior to the nominal shadow in Figure~\ref{fig:unblurred_sgra_03} is no longer evident due to blurring, its effect is to make the shadows eccentric in the tilted cases. \label{fig:blurred_sgra_03}}
\end{figure*}

The top panels in Figure~\ref{fig:blurred_sgra_03} show the results of this blurring process on five snapshots from the aligned simulation, equally spaced in time from $t = 8000$ to $t = 10{,}000$ (a span of approximately $11\ \hours$ for Sgr~A*). The bottom panels show the corresponding snapshots from the tilted simulation. By eye, adding tilt appears to make the ring brightness slightly less symmetric and to make the shadow less circular.

We use the following procedure to quantify the shapes of the rings and shadows in the blurred images. From the image center we resample intensity onto $128$ radial rays, each with $128$ sample points. A ridgeline is found, consisting of each point that is the local maximum of intensity in its ray. A new center is calculated as the centroid of the pixels contained inside this ridgeline, not weighted by intensity, and the ridgeline is recalculated from this new center, ensuring that even a feature offset from the image center is uniformly sampled in angle. This ridgeline is taken to be the ring. We define a ring ``roughness'' $\rrring$ to be the standard deviation of the set of distances from the new center to the ridgeline, divided by the mean of the set. A perfect circle would have a roughness of $0$. Our procedure so far is similar to that described in \citet[\citetalias{EHT2019d}]{EHT2019d}, their \S9.1, for defining the ring.

Next, we define the ``shadow'' to be the dimmest quartile of pixels inside the ring. Taking the zeroth and first moments of this set of pixels, again not weighted by intensity, yields the size $S$ and center $(\bar{x}^1, \bar{x}^2)$ of the shadow:
\begin{subequations} \begin{align}
  S & = \sum_\mathrm{shadow} 1, \\
  \bar{x}^i & = \frac{1}{S} \sum_\mathrm{shadow} x^i.
\end{align} \end{subequations}
A shadow contour can be defined by again sampling along rays emanating from this central point, this time finding the radius where the intensity crosses the aforementioned quartile value. This boundary also has an associated roughness value $\rrshadow$.

We are particularly interested in the shape of the shadow, so we measure it in another way. Define the second moments
\begin{equation}
  M_{ij} = \frac{4}{S} \sum_\mathrm{shadow} (x^i - \bar{x}^i) (x^j - \bar{x}^j).
\end{equation}
Arranging the second moments into a matrix, the eigenvalues can be taken to be the squares of the semimajor and semiminor axes of an ellipse, with the corresponding eigenvectors indicating the axes' orientations.\footnote{Ellipses are also generated in this way by \citet{Shiokawa2013} in the context of images of tilted disks, but there the focus is on fitting the small region of peak brightness.}

\begin{figure*}
  \centering
  \includegraphics{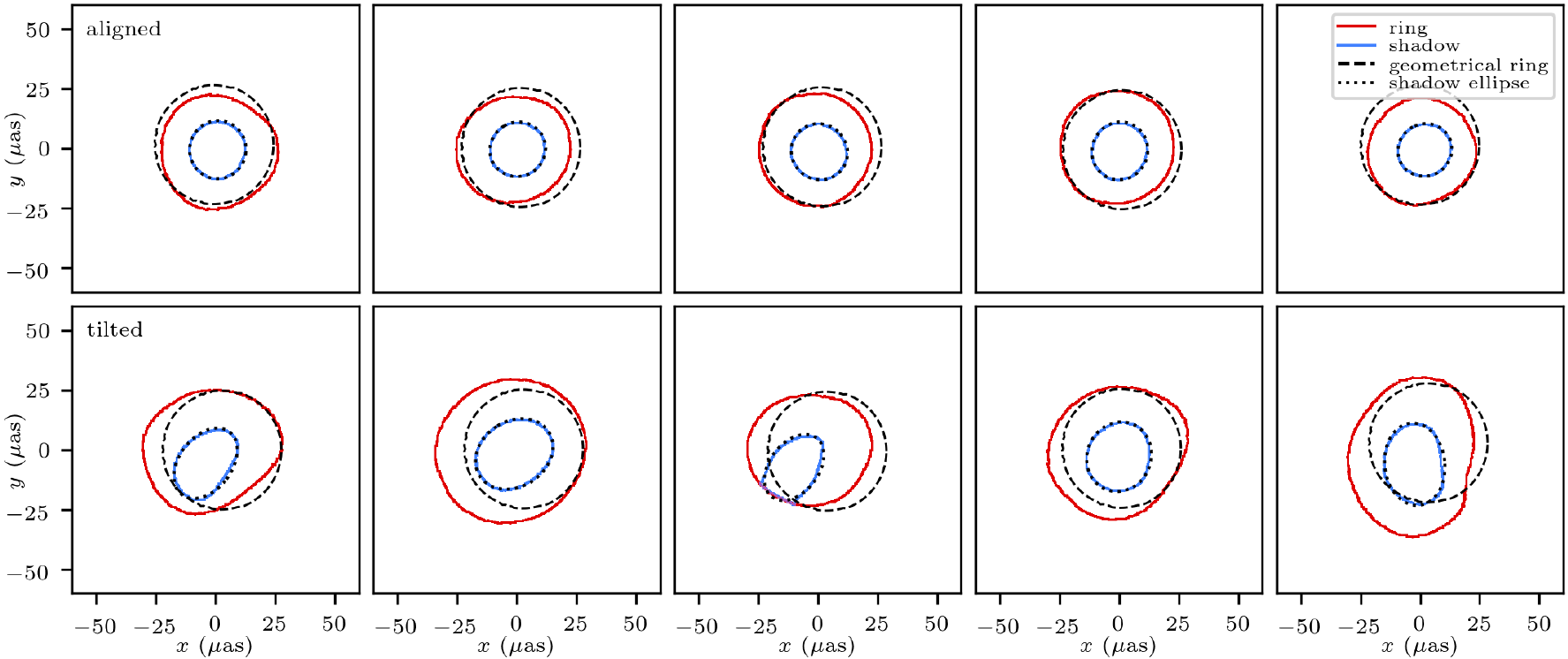}
  \caption{Contours for the ring (red), shadow (blue), geometrical ring (black dashed), and best-fit shadow ellipse (black dotted) for the aligned (top) and tilted (bottom) simulations modeling Sgr~A* face-on. The snapshots used are those shown in Figure~\ref{fig:blurred_sgra_03}. \label{fig:contours_sgra_03}}
\end{figure*}

Figure~\ref{fig:contours_sgra_03} shows the contours for the rings and shadows corresponding to the snapshots shown in Figure~\ref{fig:blurred_sgra_03}. The contours match what can be seen by eye in the blurred images. We also plot the ellipses obtained from the second moments, and they are generally good fits to the shadow regions.

Using $21$ blurred images over the span of time from $t = 8000$ to $t = 10{,}000$ (again, roughly $11\ \hours$ for Sgr~A*), we calculate the roughness parameter for the ring to be $\rrring = 0.027 \pm 0.010$ in the aligned case and $\rrring = 0.052 \pm 0.019$ in the tilted case. The reported numbers are the mean plus or minus the standard deviation over the set of snapshots. The tilted rings are slightly more uneven, but only by about one standard deviation.

The shadow roughness parameters for the same set of snapshots are $\rrshadow = 0.021 \pm 0.011$ and $\rrshadow = 0.136 \pm 0.055$ for the aligned and tilted cases, respectively. Here there is a significant difference:\ the shadows in the tilted images are distinctly noncircular. This same result is seen when examining the eccentricities of the best-fit ellipses to the shadows, which are measured to be $\eshadow = 0.291 \pm 0.085$ for the aligned images and $\eshadow = 0.69 \pm 0.11$ for the tilted images. By looking at the shape of the shadow inside the ring, rather than the ring's ridgeline itself, a difference between aligned and tilted accretion flows can be revealed.

The orientations of the major axes of the ellipses fitting the aligned shadows change drastically from one snapshot to the next (a separation of $17\ \minutes$). This is not surprising given the circular nature of these shadows. For the shadows in tilted flows, however, these orientations change slowly in time. The average magnitude by which they change from one snapshot to the next is only $13^\circ$.

\begin{figure}
  \centering
  \includegraphics{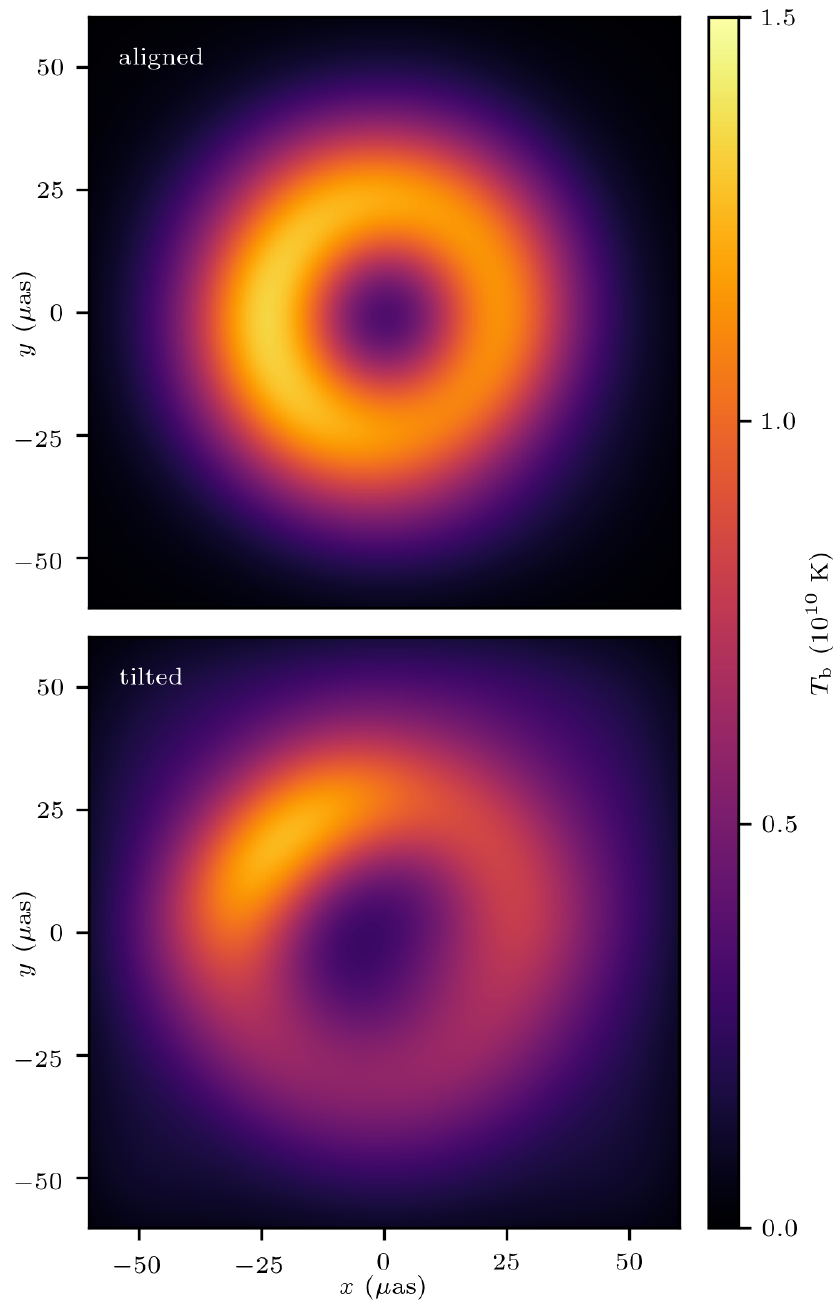}
  \caption{Blurred, time-averaged images modeled on Sgr~A* with a nearly face-on viewing angle for the aligned (top) and tilted (bottom) simulations. The camera position and orientation is the same as for the non-time-averaged images in Figure~\ref{fig:blurred_sgra_03}, and the time averaging spans $11\ \hours$. Even with this averaging, the shadow remains distinctly eccentric in the tilted case. \label{fig:blurred_sgra_03_ave}}
\end{figure}

\begin{figure}
  \centering
  \includegraphics{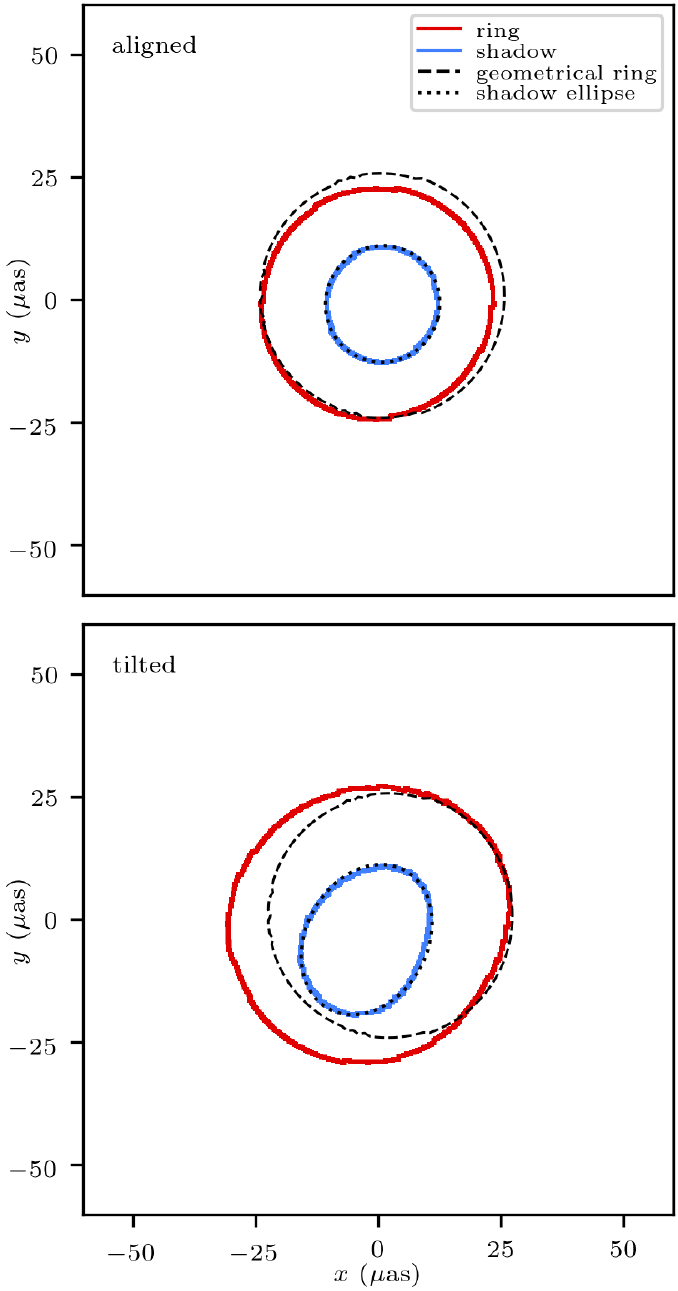}
  \caption{Contours for the ring (red), shadow (blue), geometrical ring (black dashed), and best-fit shadow ellipse (black dotted) for the aligned (top) and tilted (bottom) time-averaged images modeling Sgr~A* viewed face-on. The images used are those shown in Figure~\ref{fig:blurred_sgra_03_ave}. As is clear in Figure~\ref{fig:blurred_sgra_03_ave}, the time-averaged shadow is noticeably eccentric in the tilted simulation. \label{fig:contours_sgra_03_ave}}
\end{figure}

Given the potential rapid variability of Sgr~A* relative to the duration of EHT observations ($GM/c^3 = 20\ \seconds$), we repeat the above analyses on a time-averaged image. That is, we first apply \grtrans{} to $21$ snapshots as before, then average the resulting images in time, then blur the averaged image, and finally measure the ring and shadow properties for the single blurred image. The blurred images are shown in Figure~\ref{fig:blurred_sgra_03_ave}, and the corresponding contours are highlighted in Figure~\ref{fig:contours_sgra_03_ave}. The ring roughness is $\rrring = 0.0147$ in the aligned case and $\rrring = 0.0304$ in the tilted case. The shadow again shows a greater difference:\ $\rrshadow$ is $0.0203$ and $0.0998$ in the aligned and tilted cases, respectively, and $\eshadow$ is $0.209$ and $0.641$, respectively.

\subsection{Sgr~A* Viewed from an Angle}
\label{sec:analysis:sgra_45}

\begin{figure}
  \centering
  \includegraphics{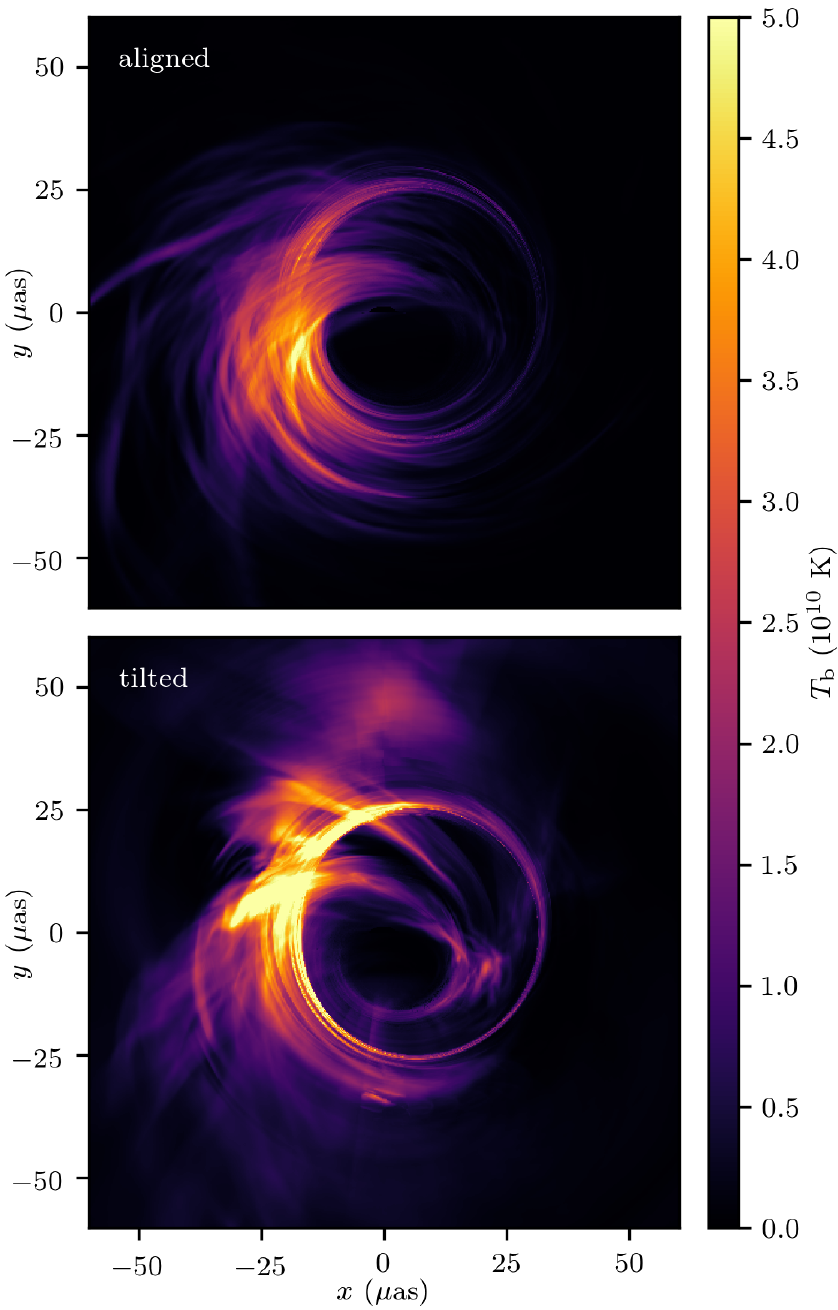}
  \caption{Images created from a snapshot from each simulation, modeled on Sgr~A* assuming an inclined viewing angle. The accretion flow is clockwise in both images, with the southern spin axis pointing toward the camera $45^\circ$ off the line of sight (pointing down when projected onto the image plane). Complex morphology is present in both images. \label{fig:unblurred_sgra_45}}
\end{figure}

Given the present uncertainty in the orientation of Sgr~A*, we consider another viewing angle:\ $45^\circ$ off the black hole spin axis. Figure~\ref{fig:unblurred_sgra_45} shows high-resolution images created from a single snapshot in both the aligned and tilted cases. Unlike in the face-on case, both images are similarly complex and there are no immediately distinguishable features.

\begin{figure}
  \centering
  \includegraphics{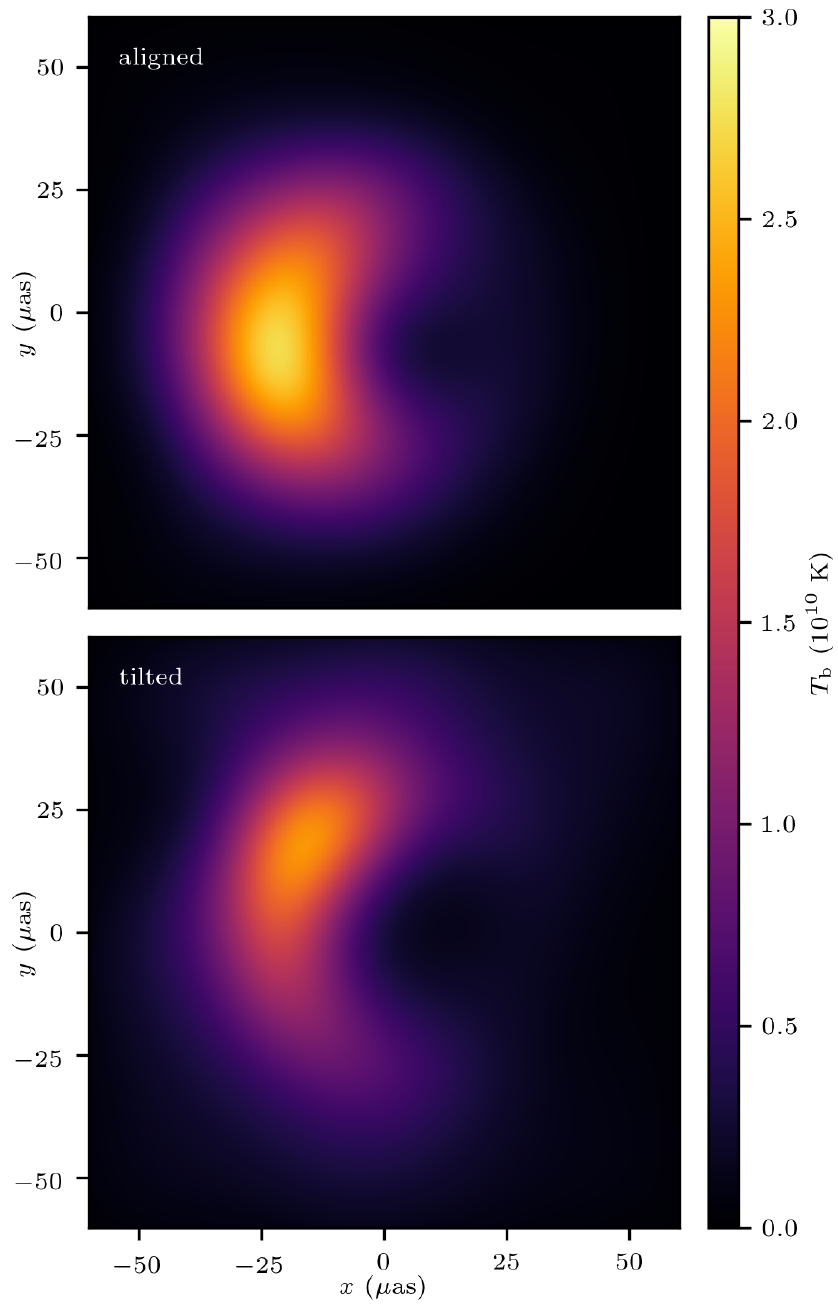}
  \caption{Blurred, time-averaged images modeled on Sgr~A* with an inclined viewing angle for the aligned (top) and tilted (bottom) simulations. The camera position and orientation is the same as for the unblurred snapshots in Figure~\ref{fig:unblurred_sgra_45}. The time averaging spans $11\ \hours$. Both cases display a crescent, but the image is more asymmetric in the tilted case, and the shadow is significantly eccentric. \label{fig:blurred_sgra_45_ave}}
\end{figure}

As in \S\ref{sec:analysis:sgra_03}, we take the blurred, time-averaged image as an appropriate proxy for EHT data. Time averaging the same $21$ snapshots spanning $t = 8000$ to $t = 10{,}000$ ($11\ \hours$ given the mass $M = 4.152 \times 10^{6}\ \msun$) and applying a $20\ \muas$ Gaussian filter, we obtain the images shown in Figure~\ref{fig:blurred_sgra_45_ave}. Images from this inclined viewing angle do not have as well-defined shadows inside rings as in the face-on case when blurred to match the EHT resolution. Still, the shadow can be seen to be significantly eccentric in the tilted image, as with the face-on viewing angle discussed in \S\ref{sec:analysis:sgra_03}. The images display a distinct crescent morphology. Importantly, the crescent is symmetric in the aligned case, whereas in the tilted case the southern tail extends further from the brightness peak.

\begin{figure}
  \centering
  \includegraphics{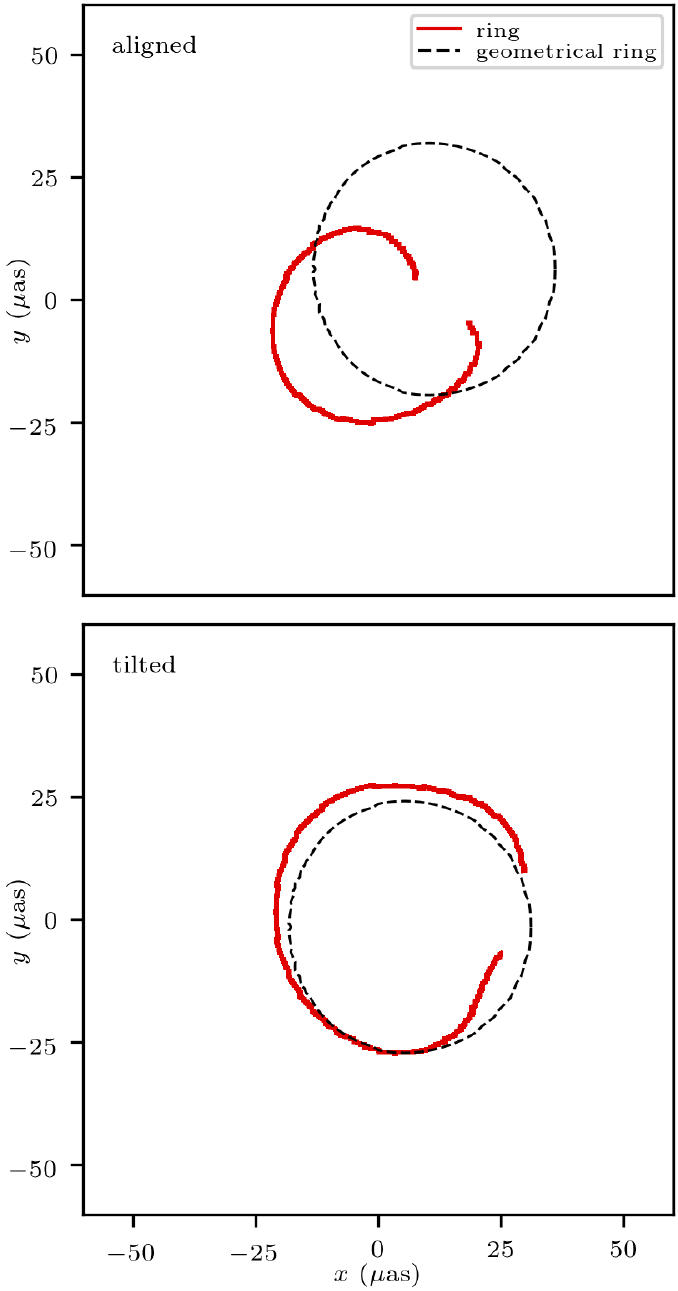}
  \caption{Ring contour (red) and geometrical ring (black dashed) for the aligned (top) and tilted (bottom) time-averaged images modeling Sgr~A* viewed at an angle. The images used are those shown in Figure~\ref{fig:blurred_sgra_45_ave}. The ring has a radius approximately $30\%$ larger in the tilted case. \label{fig:contours_sgra_45_ave}}
\end{figure}

Using the blurred images, we can define a ridgeline as before, though in some directions the algorithm will fail to find a well-defined intensity maximum. We can still define parts of a ring as shown in Figure~\ref{fig:contours_sgra_45_ave}. From these plots, it is apparent that the tilted image has a larger ridgeline. The mean distance from the ring center to the ridgeline is $\rring = 19.8\ \muas$ in the aligned case and $\rring = 26.2\ \muas$ in the tilted case.

\begin{figure}
  \centering
  \includegraphics{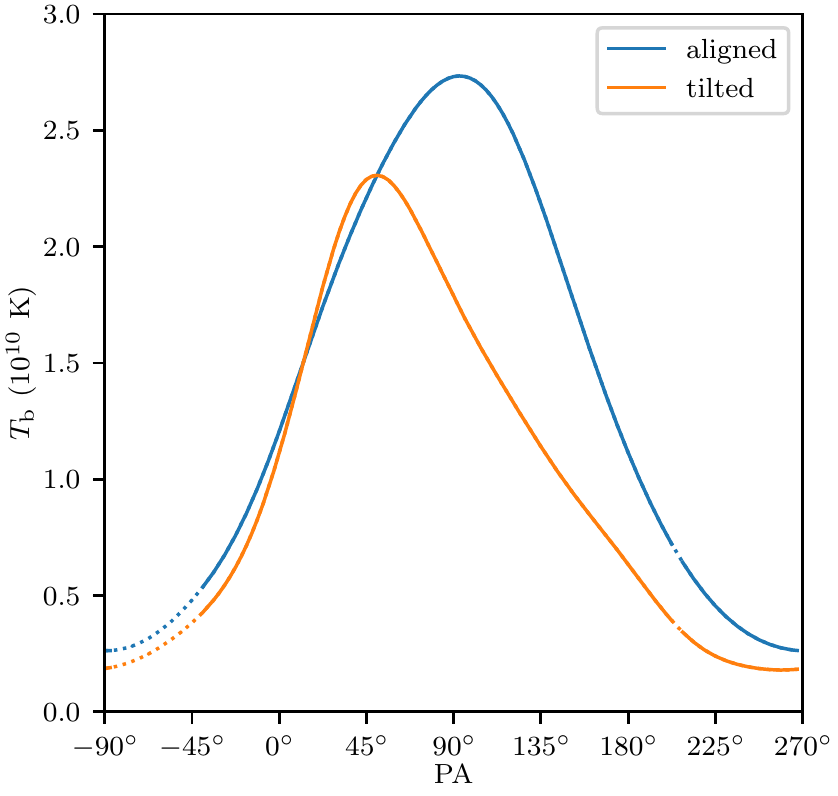}
  \caption{Ridgeline brightness temperatures for the blurred, time-averaged images from the aligned and tilted simulations, modeled to match Sgr~A* as seen at a $45^\circ$ inclination. The lines are solid where ridgelines are well defined;\ they are dashed where linear interpolation of radius as a function of position angle is used to locate them. The peak is symmetric in the aligned case and skewed in the tilted case. \label{fig:ridgeline_sgra_45_ave}}
\end{figure}

In order to sample intensity along the ridgeline, we extend the procedure:\ whenever a sample location is needed but no ridgeline can be found, the distance of the ridgeline from the central point is linearly interpolated in angle from the nearest angles in either direction for which a local maximum exists. We can then measure intensity along a closed arc. The results, plotted as a function of position angle, are shown in Figure~\ref{fig:ridgeline_sgra_45_ave}. Here again the skewness in the ridgeline intensity in the tilted case is apparent.

\subsection{Ring Size and Structure in M87}
\label{sec:analysis:m87}

\begin{figure}
  \centering
  \includegraphics{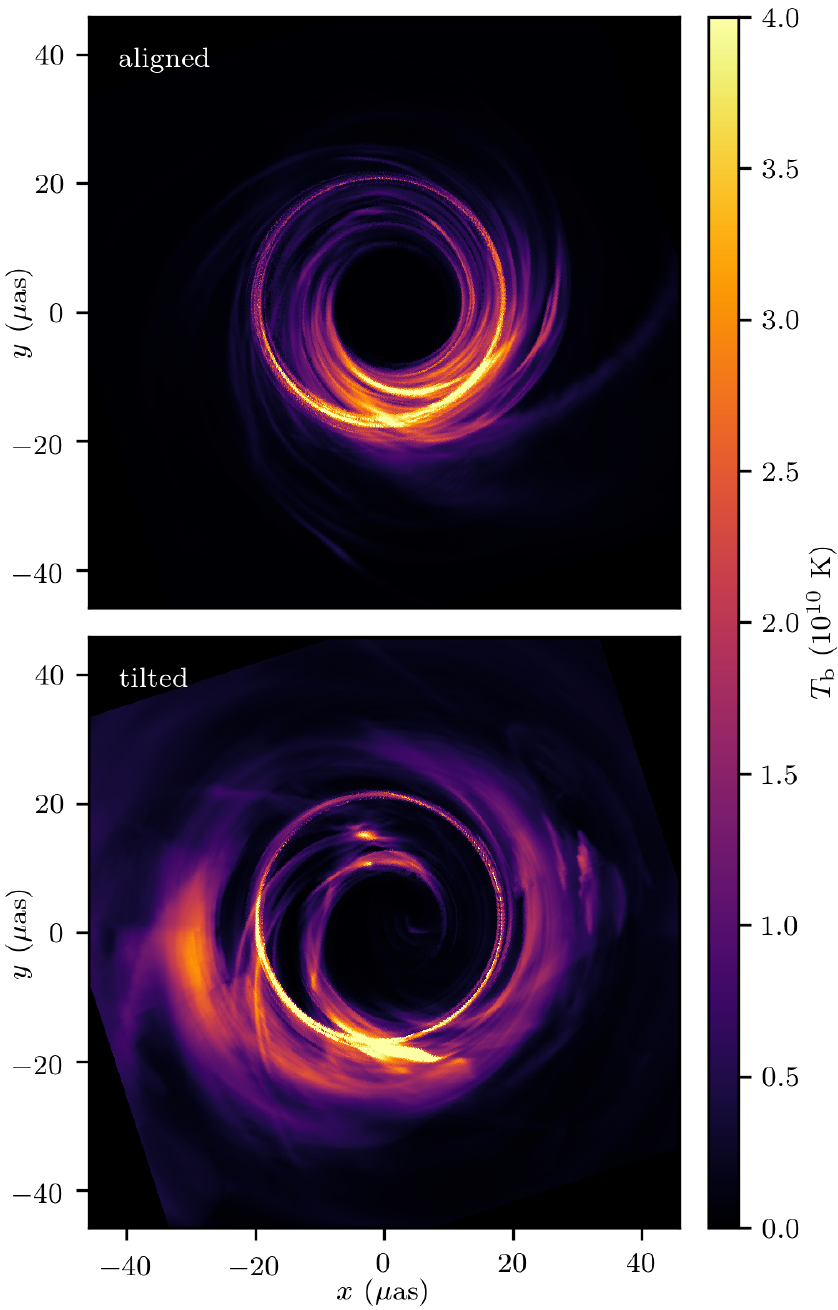}
  \caption{Images from aligned and tilted simulations modeled on M87. The accretion flow is clockwise in both images, with the southern spin axis pointing toward the camera $17^\circ$ off the line of sight (pointing to the right and slightly up, at a position angle of $288^\circ$, when projected onto the image plane). Neither case is axisymmetric, but only the tilted case has the distinct spiral shock detached from the rest of the emission near the center of the image. \label{fig:unblurred_m87}}
\end{figure}

The same two simulations can be used to model emission from M87, changing the black hole mass, distance, and average flux parameters when ray tracing. Here we do not have as much freedom to choose the viewing angle inclination $\theta_0$, given the observed large-scale jet oriented $17^\circ$ off our line of sight. That is, we assume the large-scale jet is aligned with the black hole spin axis, though there are some simulations that suggest this is not the case \citep{Liska2018}. Our choice produces images---both aligned and tilted---broadly consistent with M87 observations, with more emission coming from the southern part of the image. Moreover, changing the viewing angle to align the disk normal with the observed jet would not qualitatively change the fact that tilted flows differ from aligned ones by containing intrinsically nonaxisymmetric structure. For example, a high-resolution image from each simulation is shown in Figure~\ref{fig:unblurred_m87}, where even the aligned case now has non-axisymmetric structure. Still, the tilted case displays a qualitative difference by having a prominent spiral shock penetrate the shadow.

\begin{figure*}
  \centering
  \includegraphics{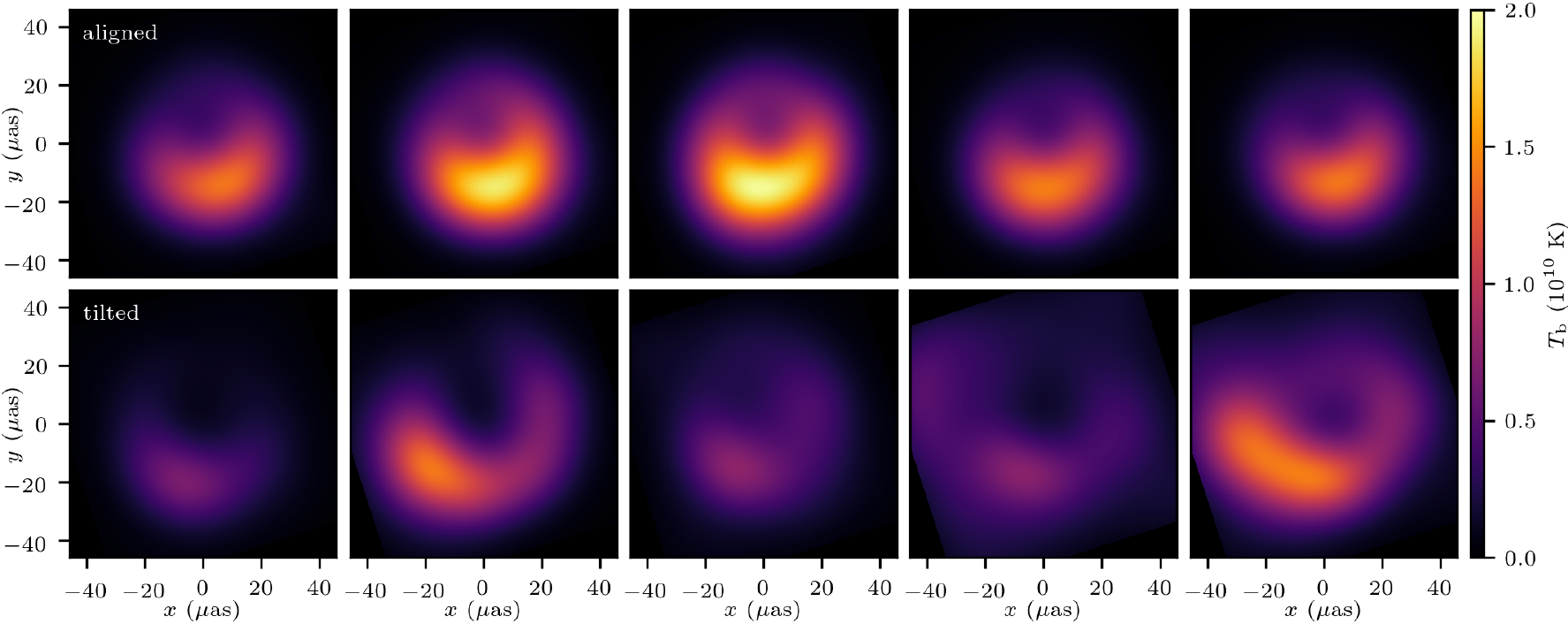}
  \caption{Blurred images modeled on M87 as seen from Earth for the aligned (top) and tilted (bottom) simulations at five different times (separated by a total of $2.0\ \years$). The camera position and orientation is the same as in Figure~\ref{fig:unblurred_m87}. Compared to the face-on images of Sgr~A* in Figure~\ref{fig:blurred_sgra_03}, the shadows here are less well defined. \label{fig:blurred_m87}}
\end{figure*}

Figure~\ref{fig:blurred_m87} shows five blurred images taken from each simulation, equally spaced over a timespan of $8000 \leq t \leq 10{,}000$ (a $2.0\ \years$ range for M87's mass). As with the inclined viewing angle for Sgr~A*, we often find the intensity ridgeline does not completely enclose the shadow;\ that is, moving outward from the shadow center the intensity sometimes monotonically decreases. Thus the simple procedure we employed for face-on Sgr~A* images fails to define a shadow here. We note this may be an artifact of our simple smoothing prescription;\ there is a ridgeline in the unblurred image, and a smaller smoothing kernel keeps it intact.

\begin{figure*}
  \centering
  \includegraphics{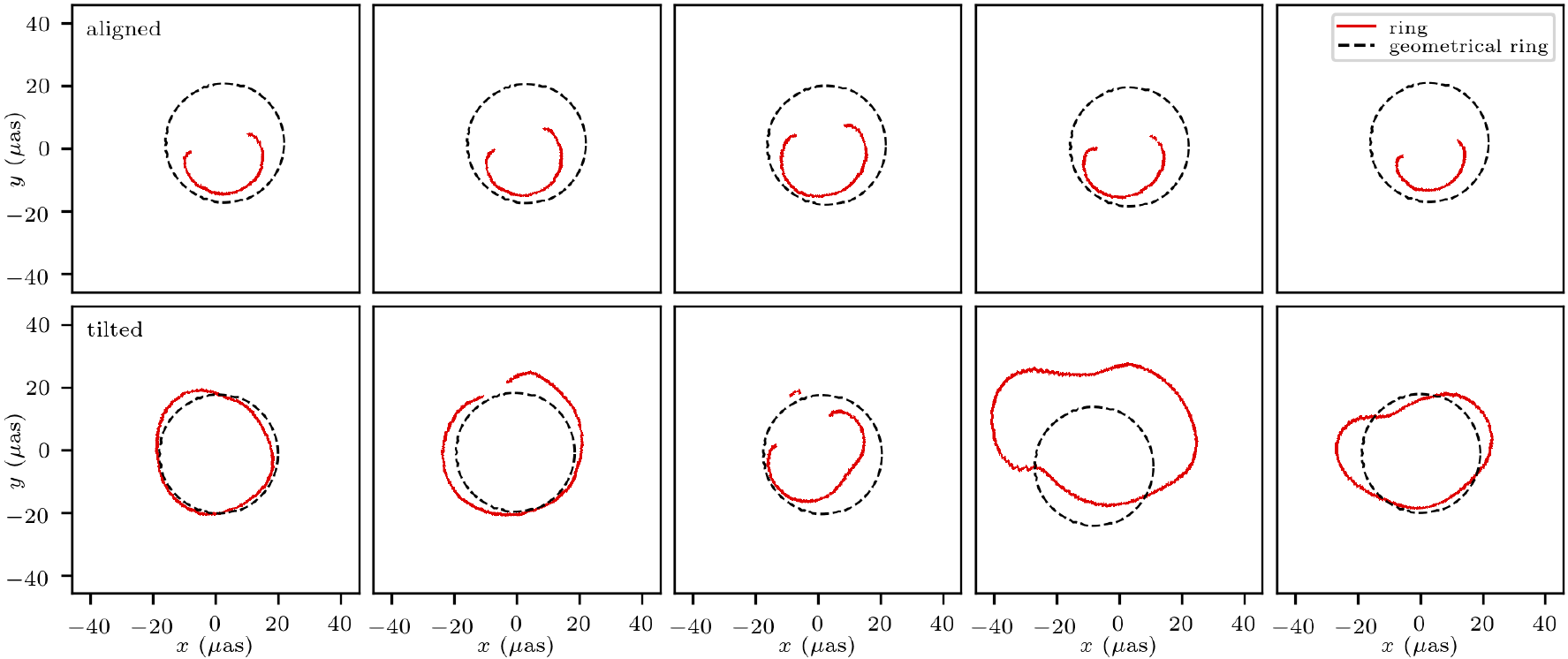}
  \caption{Ring contours (red) and geometrical rings (black dashed) for the aligned (top) and tilted (bottom) simulations modeled on M87. The snapshots used are those shown in Figure~\ref{fig:blurred_m87}. The line breaks wherever there is no local maximum and the notion of a ring becomes ambiguous. \label{fig:contours_m87}}
\end{figure*}

Figure~\ref{fig:contours_m87} shows the ridgelines corresponding to the blurred snapshots of Figure~\ref{fig:blurred_m87}. The lines break where there is no local maximum in the given direction. Using just the parts of the ridgeline that do not require interpolation, the calculated roughness parameter is $\rrring = 0.048 \pm 0.012$ in the aligned images and $\rrring = 0.083 \pm 0.048$ in the tilted images. The reported numbers are the mean plus or minus the standard deviation over $21$ snapshots.

Even though the ring and shadow are somewhat less well defined than for the face-on viewing angle, there are two immediately apparent differences between the set of aligned images and that of tilted images. First, the latter has ridgelines that are located further from the ring center in all directions. As in \S\ref{sec:analysis:sgra_45}, we quantify this by measuring a mean distance from ring center to ridgeline, averaging over all rays originating from the ring center. We then take the mean and standard deviation over the set of snapshots. In the aligned case, the average ridgeline radius is $\rring = 12.47 \pm 0.75\ \muas$;\ in the tilted case, it is $\rring = 22.0 \pm 3.4\ \muas$.

\begin{figure}
  \centering
  \includegraphics{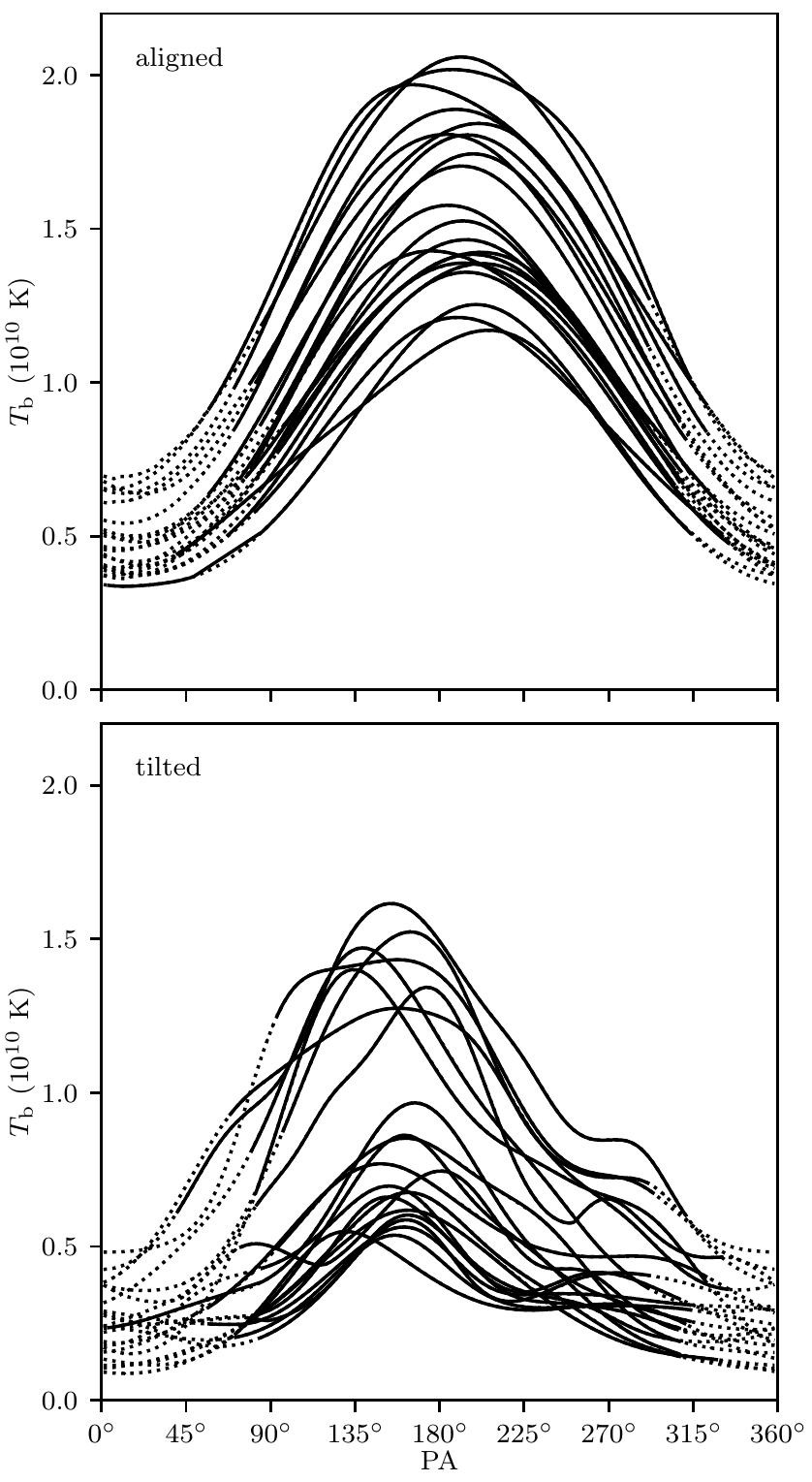}
  \caption{Ridgeline brightness temperatures for $21$ snapshots from the aligned (top) and tilted (bottom) simulations, modeled to match M87. The lines are solid where ridgelines are well defined;\ they are dashed where linear interpolation of radius as a function of position angle is used to locate them (that is, where the solid red lines in Figure~\ref{fig:contours_m87} are broken). Images of the tilted simulation are clumpier and show more local maxima in these plots. \label{fig:ridgeline_m87}}
\end{figure}

The other difference is in the regularity of the blurred aligned images relative to the tilted ones. The latter often have multiple distinct bright locations around the ring. To illustrate this better, we walk along the ridgelines and note the brightness temperature as a function of position angle. Figure~\ref{fig:ridgeline_m87} shows the resulting ring intensities for $21$ snapshots in both cases, with dotted lines denoting where interpolation is used to define a sampling location. All aligned ridgelines have a single peak, reflecting the fact that each image consists of a well-defined crescent. On the other hand, most tilted ridgelines have two or three local maxima, reflecting the clumpy nature of the images.

\section{Implications and Discussion}
\label{sec:implications}

Misalignment of infalling matter's angular momentum with that of a rapidly spinning black hole can have a significant impact on the dynamics of the accretion flow. A priori we expect such tilt to be common in low-luminosity AGN and thus to have observable consequences in resolved images as produced by the EHT. We have explored and begun to quantify some of these consequences in SANE (standard and normal evolution, as opposed to magnetically arrested) models of Sgr~A* and M87, enabling observations of these systems to provide evidence either for or against the presence of tilt.

A summary of parameters measured from our models is given in Table~\ref{tab:summary}. For completeness, we include values for $\rring$, $\rrring$, $\rrshadow$, and $\eshadow$ for all cases considered above. The average radii of the geometrical rings (the dashed lines in Figures~\ref{fig:contours_sgra_03}, \ref{fig:contours_sgra_03_ave}, \ref{fig:contours_sgra_45_ave}, and~\ref{fig:contours_m87}) are included for comparison. In the cases with inclined viewing angles, the procedure given in \S\ref{sec:analysis:sgra_03} for defining the shadow fails. However we can proceed by reducing the Gaussian blurring full width at half-maximum from $20\ \muas$ to $15\ \muas$.

\begin{deluxetable*}{lCCCC}
  \tablecaption{Summary of model parameters. \label{tab:summary}}
  \tablehead{\nocolhead{} & \colhead{Sgr~A* face-on, snapshots\tablenotemark{\footnotesize a}} & \colhead{Sgr~A* face-on, averaged} & \colhead{Sgr~A* $45^\circ$, averaged} & \colhead{M87, snapshots\tablenotemark{\footnotesize b}}}
  \startdata
  \sidehead{$\rring$}
  Geometrical\tablenotemark{\footnotesize c} ($\muas$) & 24.9\phn \phm{\pm 0.00} & 24.9   & 25.1   & 19.0\phn \phm{\pm 0.00} \\
  Aligned ($\muas$)                                    & 23.50 \pm 0.58          & 23.4   & 19.8   & 12.47 \pm 0.75          \\
  Tilted  ($\muas$)                                    & 28.0\phn \pm 1.8\phn    & 28.2   & 26.2   & 22.0\phn \pm 3.4\phn    \\[0.8ex]
  Tilted$/$aligned                                     & 1.191 \pm 0.080         & 1.20   & 1.32   & 1.76 \pm 0.29           \\
  \sidehead{$\rrring$}
  Aligned                                              & 0.027 \pm 0.010         & 0.0147 & 0.0716 & 0.048 \pm 0.012         \\
  Tilted                                               & 0.052 \pm 0.019         & 0.0304 & 0.0615 & 0.083 \pm 0.048         \\[0.8ex]
  Tilted$/$aligned                                     & 2.0 \pm 1.0             & 2.07   & 0.859  & 1.7 \pm 1.1             \\
  \sidehead{$\rrshadow$\tablenotemark{\footnotesize d}}
  Aligned                                              & 0.021 \pm 0.011         & 0.0203 & 0.138  & 0.107 \pm 0.028         \\
  Tilted                                               & 0.136 \pm 0.055         & 0.0998 & 0.113  & 0.135 \pm 0.047         \\[0.8ex]
  Tilted$/$aligned                                     & 6.6 \pm 4.4             & 4.92   & 0.817  & 1.26 \pm 0.55           \\
  \sidehead{$\eshadow$\tablenotemark{\footnotesize d}}
  Aligned                                              & 0.291 \pm 0.085         & 0.209  & 0.677  & 0.52 \pm 0.13           \\
  Tilted                                               & 0.69\phn \pm 0.11\phn   & 0.641  & 0.653  & 0.67 \pm 0.11           \\[0.8ex]
  Tilted$/$aligned                                     & 2.36 \pm 0.78           & 3.06   & 0.964  & 1.28 \pm 0.39
  \enddata
  \tablenotetext{a}{Uncertainties are standard deviations from $21$ snapshots over $11\ \hours$.}
  \tablenotetext{b}{Uncertainties are standard deviations from $18$ snapshots over $2.0\ \years$.}
  \tablenotetext{c}{Average radius of the boundary between rays that trace back through the horizon and those that do not.}
  \tablenotetext{d}{Calculated as described in \S\ref{sec:analysis:sgra_03} for the face-on Sgr~A* models, using a $20\ \muas$ blur. For Sgr~A* at an inclined viewing angle and for M87, the same procedure is applied but with a $15\ \muas$ blur in order to obtain well-defined shadows.}
\end{deluxetable*}

In the simple case of viewing a black hole along its spin axis, tilt will break axisymmetry. We consider this viewing angle for Sgr~A*. Though the orientation is likely to be different in nature, this model proves instructive. The foreground standing shock that develops in a tilted flow (tilted by $24^\circ$ in our model) results in local heating, which in turn induces local brightening with the standard electron temperature models. This bright feature approaches small radii in the image. When blurred to a resolution approximating EHT observations, the result is an eccentric shadow.

Our roughness measure $\rrshadow$, which quantifies departure from uniform circularity (a value of $0$ is circular), is significantly higher for the tilted case than the aligned case, $0.136 \pm 0.055$ instead of $0.021 \pm 0.011$ (dispersion reflecting time variability). Fitting ellipses and measuring eccentricity $\eshadow$ shows the same trend, $0.69 \pm 0.11$ instead of $0.291 \pm 0.085$. This is in contrast to the ring itself, whose roughness only increases to $0.052 \pm 0.019$ from $0.027 \pm 0.010$. These same trends hold when analyzing a time-averaged image.

Our ring roughness parameter is defined in a manner similar to the measure of circularity $\sigma_\mathrm{d} / d$ given in \citetalias{EHT2019d} (equations~\extref{18} and~\extref{19}) and plotted in \citetalias{EHT2019f} (Figure~18):\ $2 \rrring \approx \sigma_\mathrm{d} / d$. We propose, however, that the power to discriminate between aligned and tilted disks comes more from the shadow properties $\rrshadow$ or $\eshadow$ than from the properties of the ring at peak surface brightness such as $\rrring$, based on only the former and not the latter displaying strong statistically significant differences between the aligned and tilted cases we consider.

While the photon ring proper (excluding direct emission) might only become significantly noncircular with modifications to GR \citep{Johannsen2010}, the observed ring of light can obtain noncircular characteristics by merely having the accretion flow be misaligned. Constraining deviations from GR may thus require understanding the tilt of observed systems.

When the same system, whether aligned or tilted, is viewed at a much greater inclination ($45^\circ$, greater than the tilt angle), the morphology of single snapshots can no longer be captured by a simple shape. Still, time averaging leads to a clear difference between aligned and tilted flows. The radius of the ring is $30\%$ larger in the tilted case, $26.2\ \muas$ compared to $19.8\ \muas$. The crescent has a symmetric brightness distribution in the aligned case, but in the tilted case the brightest point is not centered (see Figures~\ref{fig:blurred_sgra_45_ave} and~\ref{fig:ridgeline_sgra_45_ave}). In addition, the time-averaged shadow is more eccentric in the tilted case (Figure~\ref{fig:blurred_sgra_45_ave}), just as for the face-on viewing angle.

Turning our attention to M87, with a viewing angle of $17^\circ$ (less than the disk tilt), the effects of tilt become in some ways more dramatic but also somewhat more difficult to describe succinctly. In this case, we turn to even simpler characterizations of the image, which still show differences. Our model of a tilted disk around M87 has a ring size $\rring$ that is $22.0 \pm 3.4\ \muas$, compared to $12.47 \pm 0.75\ \muas$ in the aligned case.

All else being equal, tilt can increase the size of the observed ring. As this size is used to infer the black hole mass-to-distance ratio $M/D$, estimates of this ratio may be systematically biased above the true value if tilt exists but is neglected. We note that our tilted model has a ring diameter of $2 \rring = 43.9 \pm 6.8\ \muas$, compared to the measured value of $41 \pm 1\ \muas$ for M87 \citepalias{EHT2019f}, meaning our tilted model produces a ring similar in size to that seen in M87 when using a mass of $6.5 \times 10^9\ \msun$.

Our aligned models of M87 are notable in how small the ring appears on the sky given the mass we assume. This is due to a large amount of direct emission appearing inside the photon ring proper. The high-spin, prograde SANE models in the EHT image library also show substantial emission inside the photon ring \citepalias[Figure~2]{EHT2019e}. Note that we fix the parameter $\rhigh = 10$, and prograde SANE models with $\rhigh$ values this low are ruled out in M87 on the basis of not producing enough jet power. That is, our single aligned model should not be taken to imply that M87, if aligned, must have a mass significantly higher than $6.5 \times 10^9\ \msun$. The EHT library considers more of parameter space, including more a priori viable models, and only in a minority of cases is there such a large amount of foreground emission inside the photon ring proper as in our model. In fact, it is precisely the high-spin, prograde SANE models in the library that lead to the largest inferred $M/D$ values, as shown by the rightmost column of distributions in Figure~8 of \citetalias{EHT2019e}. We have tried varying $\rhigh$ from $1$ to $100$ and find that ring size does not depend sensitively on this parameter, just as the EHT analysis finds large $M/D$ values in this case for $\rhigh$ from $10$ to $80$ and even $160$ (shown in the same figure).

Another manifestation of the effect of tilt is that of increased clumpiness in the image, as shown in ridgeline brightness profiles like Figure~\ref{fig:ridgeline_m87}. While the first image of M87 from EHT \citepalias{EHT2019a} does show multiple bright spots, this can be the result of reconstructing an image from imperfectly sampled, noisy interferometric data, a process which is not particularly well modeled by our simple $20\ \muas$ blurring. For example, Figure~10 of \citetalias{EHT2019d} shows clumpiness in reconstructions of uniform rings. Thus the irregularity of the M87 image does not itself prove there is a tilted disk, but it would be beneficial for future comparisons of raw data to models to consider tilted cases, complete with their intrinsic bright spots, and to understand what the observational requirements are to detect such clumpiness.

\begin{figure}
  \centering
  \includegraphics{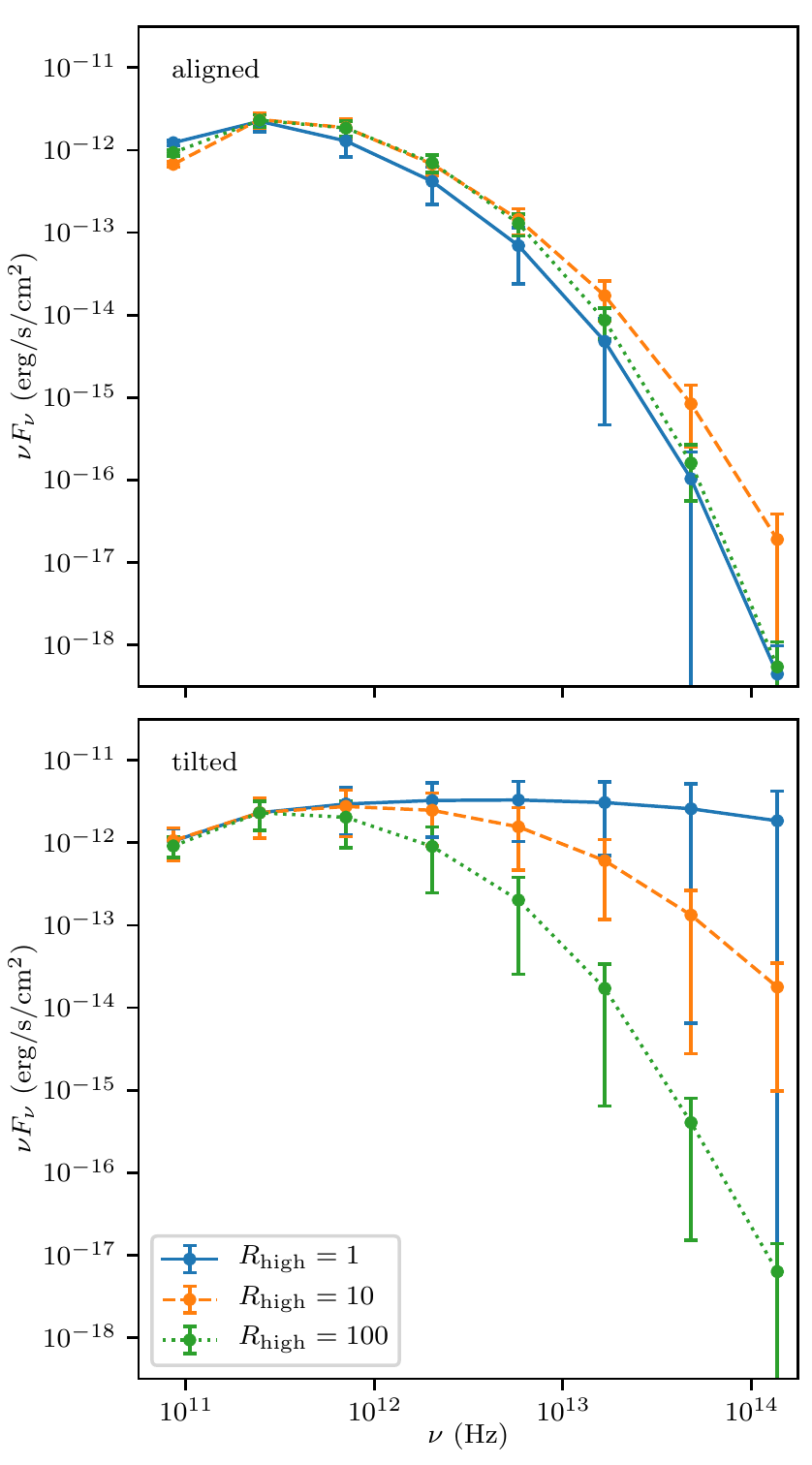}
  \caption{Spectral energy distributions of our models of M87, obtained by ray tracing at eight frequencies from the millimeter to the infrared. Error bars represent standard deviations over the $10$ snapshots used, spaced evenly over $1.8\ \years$. The results for three different $\rhigh$ parameters are shown, each independently normalized to have the same flux at $230\ \ghz$. In the tilted case, the slope of the spectrum is sensitive to $\rhigh$, which is fixed to $10$ throughout the rest of this work. \label{fig:spectra}}
\end{figure}

Despite the differences between images of aligned and tilted disks for M87, the latter are not so discrepant with existing data as to be immediately ruled out. For example, with the electron model and blurring adopted here, the morphology we find is still largely ringlike for the tilted case (see Figure~\ref{fig:blurred_m87}), in contrast to the two distinct lobes seen in \citet{Dexter2013}. We also note that the model parameters we adopt produce spectral energy distributions (SEDs) broadly consistent with those observed for M87, as reported in \citet{Prieto2016} for example. Figure~\ref{fig:spectra} shows the SEDs we obtain in the aligned and tilted cases, using three different $\rhigh$ values. Recent work by \citet{Chatterjee2020} finds that the SED of M87 is well fit by tilted models, and that these models also do well at simultaneously fitting the position angles of the large-scale jet and $230\ \ghz$ emission.

We have only examined a single tilted simulation with a single aligned comparison simulation, in order to highlight the most important qualitative differences between the two cases when it comes to horizon-scale observations. Further exploration of this additional parameter is certainly warranted, in order to answer more quantitative questions beyond the scope of this work. For example, the effects we see should grow stronger with increasing tilt, just as the shock heating grows, but the exact dependence is not determined. As tilt approaches $90^\circ$, the two-armed spiral pattern may be replaced by a different flow structure. At the same time, there should be spin and tilt angles below which we expect to see essentially the same image as produced by an untilted disk. These upper and lower cutoffs are undetermined.

Here we only consider models in the high-spin, prograde regime. In the EHT library, these differ from low-spin and retrograde models by having a large amount of direct emission inside the photon ring proper. The differences, such as in ring size, we see between aligned and tilted M87 models may not hold in other portions of parameter space.

Further, the question remains whether these results for largely incoherent magnetic fields apply to the magnetically arrested regime. This is particularly important for application to M87.

Aside from running additional GRMHD simulations to cover more of parameter space, there are lines of inquiry that can be done with the same models. We are currently investigating in more detail the effects of tilt on spectral signatures and polarization, in anticipation of further observations and analysis of such being completed in the near future.

\acknowledgments

We thank O.~Porth for suggestions for improving the manuscript, as well as A.~Spitkovsky for useful discussions about electron heating.

\strut

This research was supported in part by the National Science Foundation under grants NSF~AST~1715054 and NSF~PHY~1748958 and by a Simons Investigator award from the Simons Foundation (EQ). This work used the Extreme Science and Engineering Discovery Environment (XSEDE) Stampede2 at the Texas Advanced Computing Center through allocation AST170012.

\software{\grtrans{} \citep{Dexter2009,Dexter2016}, \athena{} \citep{White2016}}

\bibliographystyle{aasjournal}
\bibliography{references}

\end{document}